\documentclass[12pt]{article}

\usepackage{scicite}
\usepackage{times}
\usepackage{caption}
\usepackage{graphicx} 
\usepackage{amsmath}  
\usepackage{amssymb}  
\usepackage{url}      

\newenvironment{sciabstract}{%
\begin{quote} \bf}
{\end{quote}}

\title{An ultra-dense Neptune-sized planet \\
in the hot-Neptune desert} 

\author
{L. Naponiello,$^{1,2,3\,\ast}$
L. Mancini,$^{1,3,4}$   
A. Sozzetti,$^{3}$      
A.~S. Bonomo,$^{3}$     
A. Morbidelli,$^{5}$    
\\
J. Dou,$^{6}$           
L. Zeng,$^{7,8}$        
Z.~M. Leinhardt,$^{6}$  
K. Biazzo,$^{9}$       
P. Cubillos,$^{3,10}$    
M. Pinamonti,$^{3}$     
\\
D. Locci,$^{11}$        
A. Maggio,$^{11}$       
M. Damasso,$^{3}$       
A.~F. Lanza,$^{12}$     
J.~J. Lissauer,$^{13,14}$  
\\ [4pt]
A. Bignamini,$^{15}$   
W. Boschin,$^{16,17,18}$ 
L.~G. Bouma,$^{19,20}$   
P.~J. Carter,$^{6}$    
D.~R. Ciardi,$^{20}$   
\\
K.~A. Collins,$^{7}$   
R. Cosentino,$^{16}$   
I. Crossfield,$^{21}$
S. Desidera,$^{22}$     
X. Dumusque,$^{23}$     
\\
A.~F.~M. Fiorenzano,$^{16}$ 
A. Fukui,$^{35,17}$          
P. Giacobbe,$^{3}$      
C.~L. Gnilka,$^{13,20}$ 
A. Ghedina,$^{16}$      
\\
E. Gonzales,$^{24}$
G. Guilluy,$^{3}$       
A. Harutyunyan,$^{16}$  
S.~B. Howell,$^{13}$    
J.~M. Jenkins,$^{13}$   
\\
M.~B. Lund,$^{20}$      
E.~L.~N. Jensen,$^{25}$ 
J.~F. Kielkopf,$^{26}$  
K.~V. Lester,$^{13}$    
\\
L. Malavolta,$^{22,27}$ 
A.~W. Mann,$^{28}$     
R.~A. Matson,$^{29}$   
E.~C. Matthews,$^{4}$    
\\
D. Nardiello,$^{22}$   
N. Narita,$^{36,37,17}$           
E. Pace,$^{2}$         
I. Pagano,$^{12}$      
E. Palle,$^{17,18}$    
M. Pedani,$^{16}$      
\\
S. Seager,$^{8,30,31}$ 
J.~E. Schlieder,$^{32}$ 
R.~P. Schwarz,$^{7}$   
A. Shporer,$^{30}$     
\\
J.~D. Twicken,$^{13,33}$ 
J.~N. Winn,$^{34}$     
C. Ziegler,$^{35}$     
T. Zingales$^{22,27}$  
}

\date{}

\begin{document} 

\maketitle 

\vspace{-0.80cm}
\hspace{-0.85cm}
$^{1}$Department of Physics, University of Rome ``Tor Vergata'', Rome, Italy.
$^{2}$Department of Physics and Astronomy, University of Florence, Florence, Italy.
$^{3}$INAF, Turin Astrophysical Observatory, Pino Torinese, Italy.
$^{4}$Max Planck Institute for Astronomy, Heidelberg, Germany.
$^{5}$Laboratoire Lagrange, Universit\'{e} Cote d’Azur, CNRS, Observatoire de la C$\hat{\rm o}$te d’Azur, France.
$^{6}$University of Bristol, H. H. Wills Physics Laboratory, Bristol, UK. 
$^{7}$Center for Astrophysics \textbar \ Harvard \& Smithsonian, Cambridge, MA, USA.
$^{8}$Department of Earth and Planetary Sciences, Harvard University, Cambridge, MA, USA.
$^{9}$INAF -- Rome Astronomical Observatory, Monte Porzio Catone, Italy.
$^{10}$Space Research Institute, Austrian Academy of Sciences, Graz, Austria.
$^{11}$INAF -- Palermo Astronomical Observatory, Palermo, Italy.
$^{12}$INAF -- Catania Astrophysical Observatory, Catania, Italy.
$^{13}$NASA Ames Research Center, Moffett Field, CA, USA.
$^{14}$Department of Earth \& Planetary Sciences, Stanford University, Stanford, CA, USA
$^{15}$INAF -- Trieste Astronomical Observatory, Trieste, Italy.
$^{16}$Fundaci\'{o}n Galileo Galilei - INAF, Tenerife, Spain
$^{17}$Instituto de Astrof\'isica de Canarias (IAC), Tenerife, Spain.
$^{18}$Departamento de Astrof\'isica, Universidad de La Laguna (ULL), Tenerife, Spain.
$^{19}$51 Pegasi b Fellow
$^{20}$NASA Exoplanet Science Institute-Caltech/IPAC, Pasadena, CA, USA.
$^{21}$Department of Physics and Astronomy, University of Kansas, Lawrence, KS 66045, USA
$^{22}$INAF - Padova Astronomical Observatory, Padova, Italy.
$^{23}$Observatoire de Gene\'{e}ve, Universit\'{e} de Gene\'{e}ve, CH-1290 Versoix, Switzerland 
$^{24}$Department of Astronomy and Astrophysics, University of California, Santa Cruz, CA 95064, USA
$^{25}$Department of Physics \& Astronomy, Swarthmore College, PA, USA.
$^{26}$Department of Physics and Astronomy, University of Louisville, KY, USA.
$^{27}$Dipartimento di Fisica e Astronomia, Università degli Studi di Padova, PD, Italy
$^{28}$Department of Physics and Astronomy, The University of North Carolina, NC, USA.
$^{29}$U.\,S. Naval Observatory, Washington, D.C., USA.
$^{30}$Department of Physics and Kavli Institute for Astrophysics and Space Research, Massachusetts Institute of Technology, MA, USA.
$^{31}$Department of Aeronautics and Astronautics, MIT, Cambridge, MA, USA
$^{32}$NASA Goddard Space Flight Center, Greenbelt, MD, USA.
$^{33}$SETI Institute, Mountain View, CA, USA 
$^{34}$Department of Astrophysical Sciences, Princeton University, NJ, USA.
$^{35}$Dep. of Physics, Engineering and Astronomy, Stephen F. Austin State University, TX, USA.
$^{36}$Komaba Institute for Science, The University of Meguro, Japan
$^{37}$Astrobiology Center, Osawa, Mitaka, Japan

\hspace{-0.85cm}
{\bf $^\ast$Corresponding author: E-mail: luca.naponiello@unifi.it}

\begin{sciabstract} %
The paucity of Neptune-type planets at short orbital periods was recognised in the statistical studies of exoplanet populations and is known as `hot-Neptune desert' \cite{2016A&A...589A..75M}.
As many Neptune planets have been discovered with longer orbital periods, this dearth is not caused by observational biases \cite{2013ApJS..204...24B,2014ApJ...790..146F} but primarily by atmospheric photo-evaporation effects \cite{Owen2018,2022AJ....164..234V}.
Since then, the desert has been increasingly populated with planets \cite{2016ApJS..226....7C,2018AJ....156..277L} having a wide range of different and, in some cases, unusual characteristics \cite{2022A&A...666A.184P,2020Natur.583...39A}. 
%
%
The study of exoplanets that have experienced unconventional evolutionary processes offers a novel opportunity to enhance our current understanding of planetary formation and composition theories.
Here we report the discovery of the transiting planet TOI-1853\,b, which orbits a dwarf star every 1.24\,days. This planet has a mass of $73.5^{+4.2}_{-4.0}$ Earth masses, almost twice that of any other Neptune-sized planet known. These values place TOI-1853\,b in the middle of the Neptunian desert and imply that its mass is dominated by heavy elements. 
The remarkable properties of TOI-1853\,b could be the result of multiple proto-planets collisions or the final state of an initially high-eccentricity planet which migrated closer to its parent star.
\end{sciabstract}

\section{Main text}

TOI-1853 is a dwarf star with a $V$-band optical brightness of $12.3$ magnitudes, located $167$\,pc from the Sun \cite{methods}. It was photometrically monitored by the TESS space telescope and the analysis of its light curve \cite{methods} showed transit-like events compatible with a planet candidate, designated as TOI\,1853.01, having a short orbital period of $1.24$\,days and a Neptune-like radius. We ruled out a nearby eclipsing binary (NEB) blend as the potential source of the TOI-1853.01 detection in the wide \emph{TESS} pixels, by monitoring extra transit events with the higher angular resolution of three ground-based telescopes: MuSCAT2, ULMT and LCOGT \cite{methods}.
As part of the standard process for validating transiting exoplanets and assessing the possible contamination of companions on the derived planetary radii \cite{2015ApJ...805...16C}, we observed TOI-1853 with near-infrared adaptive optics imaging, using the NIRC2 instrument on the Keck-II telescope, and with optical speckle imaging, using the $^{\backprime}$Alopeke speckle imaging camera at Gemini North, and the High Resolution Imaging on the 4.1\,m Southern Astrophysical Research (SOAR) telescope. No nearby stars bright enough to significantly dilute the transits were detected within $0.5^{\prime \prime}$, $1^{\prime \prime}$ and $3^{\prime \prime}$ of TOI-1853 in the Gemini, Keck and SOAR observations, respectively \cite{methods}. Using Gaia DR3 data \cite{2021A&A...649A...1G} we also found that the astrometric solution is consistent with the star being single \cite{methods}.

We monitored TOI-1853 in the context of the Global Architecture of Planetary Systems programme \cite{Konig2022,Naponiello2022}, with the HARPS-N spectrograph \cite{2012SPIE.8446E..1VC} at the Telescopio Nazionale Galileo in La Palma. The HARPS-N data reduction software pipeline provided wavelength-calibrated spectra \cite{methods}, which we used to determine the stellar atmospheric properties. TOI-1853 is a K2\,V star with effective temperature $T_{\rm eff} = 4985 \pm 70$\,K, surface gravity $\log{g}=4.49\pm0.11$ dex, iron abundance [Fe/H] $=0.11\,\pm\,0.08$\,dex, and solar Fe/Si - Mg/Si ratios \cite{methods}. Furthermore, we determined a mass of $M_{\star} = 0.837 \pm 0.036$ solar masses ($M_{\odot}$), a radius of $R_{\star} = 0.808 \pm 0.009$ solar radii ($R_{\odot}$), and an advanced, though uncertain, stellar age of $7.0^{+4.6}_{-4.3}$~Gyr (Table\,1). 
We computed the Generalised Lomb-Scargle periodogram of the HARPS-N RVs and found a significant peak (with False Alarm Probability - FAP $\ll$ 0.1\%) at a frequency of $\approx0.8$\,d$^{-1}$ that matches the transit period and the phase of the planet candidate. 

To determine the main physical and orbital parameters of the system, we performed a transit and RV joint analysis \cite{methods}. Fig.\,1 displays the \emph{TESS} photometric light curve, together with the transits and the HARPS-N RV data phase-folded to the best fit period of the candidate. We measured the radius of the companion to be $3.45^{+0.13}_{-0.14}$\,Earth radii $(R_{\oplus})$, with a mass of $73.5^{+4.2}_{-4.0} \, M_{\oplus}$, thus confirming the planetary nature of TOI-1853.01, hereafter TOI-1853\,b. These values imply a bulk density of $9.8^{+1.9}_{-1.5}$ g\,cm$^{-3}$ ($\approx$\,6 times that of Neptune) and surface gravity of $g_{\rm p} = 60^{+14}_{-11}$ m\,s$^{-2}$ ($\approx$\,5.5 times that of Neptune), as detailed in Table\,1. 
Despite its short orbital period, TOI-1853\,b may not be engulfed during the remaining main-sequence lifetime of its host \cite{methods}.

The exceptional properties of TOI-1853\,b are clearly evident in comparison with the currently known exoplanet population (Fig.\,2). Objects with the same density of TOI-1853\,b are rare, typically super-Earths, 
while planets with the same mass usually have radii more than twice as large.
Furthermore, it occupies a region of the mass-orbital period space of hot planets that was previously devoid of objects, corresponding to the driest area of the hot-Neptune desert \cite{Owen2018}. 
TOI-1853\,b is twice as massive as the two runners-up with similar radius in the radius-mass diagram (Fig.\,2), i.e. the ultra-hot ($P=0.76$\,days) Neptune-sized TOI-849\,b \cite{2020Natur.583...39A} and the warm ($P=55$\,days) Neptune HD\,95338\,b \cite{Diaz2020}. 

While for HD\,95338\,b and TOI-849\,b the atmospheric mass fraction is expected to be at most $\sim5-7\%$ \cite{Kubyshkina2022} and $\sim4\%$ \cite{2020Natur.583...39A}, respectively, TOI-1853\,b is best described as a bare core of half water–half rock with no or negligible envelope, or as having at most 1\% atmospheric H/He mass fraction on top of a 99\% Earth-like rocky interior (Fig.\,2). 
The characteristic pressure of its deep interior is estimated to reach $\sim 5000$\,GPa (50 times the core-mantle boundary pressure of Earth), where most elements and their compounds are expected to metalize due to the reduced spacings of neighbouring atoms under extreme compression. TOI-1853\,b's metallic core could be surrounded by a mantle constituted by H$_2$O in a high-pressure ice phase and, possibly, also in liquid form \cite{Zeng2021}, although the properties of matter at such high central pressures are still quite uncertain and compositional mixing \cite{Bodenheimer2018,Vazan2022,Kovacevic2022,Stevenson2022} might be present rather than distinct layers as postulated by standard models \cite{Dorn2017,Zeng2019}. 

If TOI-1853\,b is a water-rich world, its upper structure could be described in terms of a hydrosphere with variable mass fractions of supercritical water on top of a mantle-like interior \cite{Mousis2020}. While existing structural models of Neptunes addressing this possibility do not encompass objects with a mass similar to that of TOI-1853\,b, atmospheric characterization measurements with the James Webb Space Telescope might be telling. For instance, by combining at least three transit observations with the NIRISS/SOSS instrument it should be possible to detect the series of H$_2$O absorption bands in the $0.9-2.8 \, \mu$m range, which might allow us to distinguish a thin, H$_2$-dominated atmosphere from an H$_2$O-dominated atmosphere \cite{methods}.

It is difficult to explain the formation of planets with such a large amount of heavy elements. Pebble accretion, which is the most efficient growth process for massive planets, shuts off when the core is massive enough to disrupt the gas disk \cite{Lambrechts2014}, and accreting solids beyond this mass requires a different process. The planetesimal isolation mass for runaway growth \cite{Safronov1972} as well post-isolation growth \cite{Lissauer1987} would require exceedingly and unrealistically high surface densities to grow a planet composed almost entirely of condensable material in situ. Thus, the growth of a planet like TOI-1853\,b by planetesimal accretion alone appears unrealistic as well.

One possibility is that a system of small planets migrated from the distant regions of the disk towards its inner edge, which could have loaded the inner disk of solid mass. After the disappearance of gas from the disk, the system became unstable, leading to several mutual collisions among the small planets and eventually forming a planet with a large mass of heavy elements, by merging. Our simulations \cite{methods} show that the formation of a planet with a large mass of heavy elements by merging of solid-rich planets is possible, even though merging into a single planet is an unlikely event. 
%
Moreover, groups of small planets carrying cumulatively $\sim 100\,M_{\oplus}$ in proto-planetary disks might be rare, although the {\it Kepler}-mission results indicate this scenario is not unrealistic \cite{2019A&A...624A..15S}.

Another possible formation scenario is based on the jumping-model \cite{Beauge2012}, in which at least three giant planets form at a few au from their parent star. After the disappearance of the disk, this system becomes unstable and suffers mutual scattering, leading to high-eccentricity orbits for the surviving planets, which then can be circularized by tidal damping at perihelion passages. 
If the inner disk initially contained a lot of solid mass, in the form of planetesimals or small proto-planets, the innermost planet would have engulfed most of it near perihelion. To test this possibility, we simulated an eccentric Jupiter-mass planet with an initial budget of 20 to $40\,M_{\oplus}$ in heavy elements and found that it can accrete an additional $30-40\,M_{\oplus}$ \cite{methods}. 
An initially massive H/He dominated giant planet TOI-1853\,b could have lost the bulk of its envelope mass due to tidal stripping \cite{Beauge2013,Owen2018} near periastron passage during the high-eccentricity migration. The planet that we see now may have survived very close passages to its parent star from early in the system's history since the host star would have shrunk to a size of less than 2 $R_{\odot}$ well before reaching the main sequence. Tidal heating of a young giant planet still hot from accretion would have further expanded its envelope, facilitating the escape of light gases from the original atmosphere \cite{Beauge2013,Owen2018,Owen2019}.

The anomalous density and location within the Neptunian desert of TOI-1853\,b present a fascinating puzzle for planet formation and evolution theories, thus making it a remarkable addition to the growing catalogue of exoplanets. Since the local merging of solid-rich proto-planets rarely develops into a single planet, while the migration scenario would have removed all objects within $\sim$1\,au, we computed the sensitivity limits of HARPS-N RVs to additional planetary companions \cite{methods}, and found that we can only exclude, with a 90\% confidence level, the presence of companions of masses $>10\, M_{\oplus}$ up to orbital periods of $\lesssim 10$\,days and with masses $>30\, M_{\oplus}$ up to $\lesssim 100$\,days. Eventually, by searching for planetary companions and investigating its possibly thin atmosphere, new data collection will help us unveil both the formation and composition of the densest Neptune-sized planet currently known.


\begin{table}
\caption{\textbf{Table 1: Stellar and planetary properties}. The errors represent the 68\% confidence intervals (one standard deviation or $\sigma$) for each value. The equilibrium temperature is estimated for a zero Bond albedo, in the assumption of uniform heat redistribution to the night side. The eccentricity upper limit is also constrained at the confidence level of $1\sigma$.}
\centering
\begin{tabular}{lcc}
\\
\textbf{Parameters} & \textbf{Unit} &   \textbf{Value}   \\
\hline \\ [-8pt]
\textit{Stellar} & &  \\
Right ascension (J2020)\dotfill &  & 14:05:50.24\\
Declination (J2020)\dotfill &  &  +16:59:32.53~~  \\
Spectral type\dotfill &  & K2.5V   \\
$V$-band magnitude\dotfill & mag & 12.276 $\pm$ 0.092   \\[2pt]
Parallax\dotfill  &mas & 6.022 $\pm$ 0.016  \\
Distance\dotfill  & pc & 166.8$^{+0.9}_{-0.5}$  \\
$v\sin{ i}$\dotfill  & km\,s$^{-1}$ & 1.3 $\pm$ 0.9  \\
$\log R^{\prime}_{\rm HK}$\dotfill & dex &  $-4.73\pm0.06$ \\
Mass, $M_{\star}$\dotfill  & $M_{\odot}$ & 0.837$^{+0.038}_{-0.034}$   \\[2pt]
Radius, $R_{\star}$\dotfill  & $R_{\odot}$ & 0.808 $\pm$ 0.009   \\ 
Luminosity, $L_{\star}$\dotfill & $L_{\odot}$ & 0.3696 $\pm$ 0.0093  \\
Effective temperature, $T_{\rm eff}$\dotfill & K & 4985 $\pm$ 70  \\
Surface gravity, $\log{g}$\dotfill & dex & $4.49\pm0.11$ \\
Iron abundance,      [Fe/H]\dotfill  & dex & $0.11 \pm 0.08$  \\
Magnesium abundance, [Mg/H]\dotfill  & dex & $0.09 \pm 0.06$  \\
Silicon abundance,   [Si/H]\dotfill  & dex & $0.14 \pm 0.06$  \\ [6pt]
\textit{Planetary} & &  \\
Orbital period, $P$\dotfill  & days  & 1.2436275$^{+0.0000027}_{-0.0000031}$ \\[2pt]
Radial velocity semi-amplitude, $K$\dotfill  & m\,s$^{-1}$ & 49.0 $\pm$ 1.2  \\
Eccentricity, $e$\dotfill &  & $<0.03$  \\
Argument of periastron, $\omega$\dotfill & degrees  & unconstrained  \\
Impact parameter, $b$\dotfill &  & $0.51^{+0.06}_{-0.07}$ \\
Reference epoch of mid-transit, $T_{0}$\dotfill & BJD$_{\rm TDB}$  & 2\,459\,690.7419 $\pm$ 0.0008  \\
Transit duration, $T_{14}$\dotfill & hours & $1.49\pm0.09$ \\
Orbital semi-major axis, $a$\dotfill & au & 0.021 $\pm$ 0.001  \\[2pt]
Mass, $M_{\rm p}$\dotfill  & $M_{\oplus}$ & 73.5 $^{+4.2}_{-4.0}$ \\ [2pt]
Radius, $R_{\rm p}$\dotfill  & $R_{\oplus}$ & 3.45$^{+0.13}_{-0.14}$  \\ [2pt]
Inclination, $i$\dotfill  & degrees & 84.9$^{+0.7}_{-0.6}$ \\ [2pt]
Density, $\rho_{\rm p}$\dotfill  & g\,cm$^{-3}$ & $9.8^{+1.9}_{-1.5}$ \\ [2pt]
Surface gravity, $g$\dotfill  & m\,s$^{-2}$ & 60$^{+14}_{-11}$  \\ [2pt]
Equilibrium temperature, $T_{\rm eq}$\dotfill  & K & 1478 $\pm$ 35 \\
\hline
\end{tabular}
\end{table}

\clearpage

\begin{figure}
\centering
\includegraphics[width=1\textwidth]{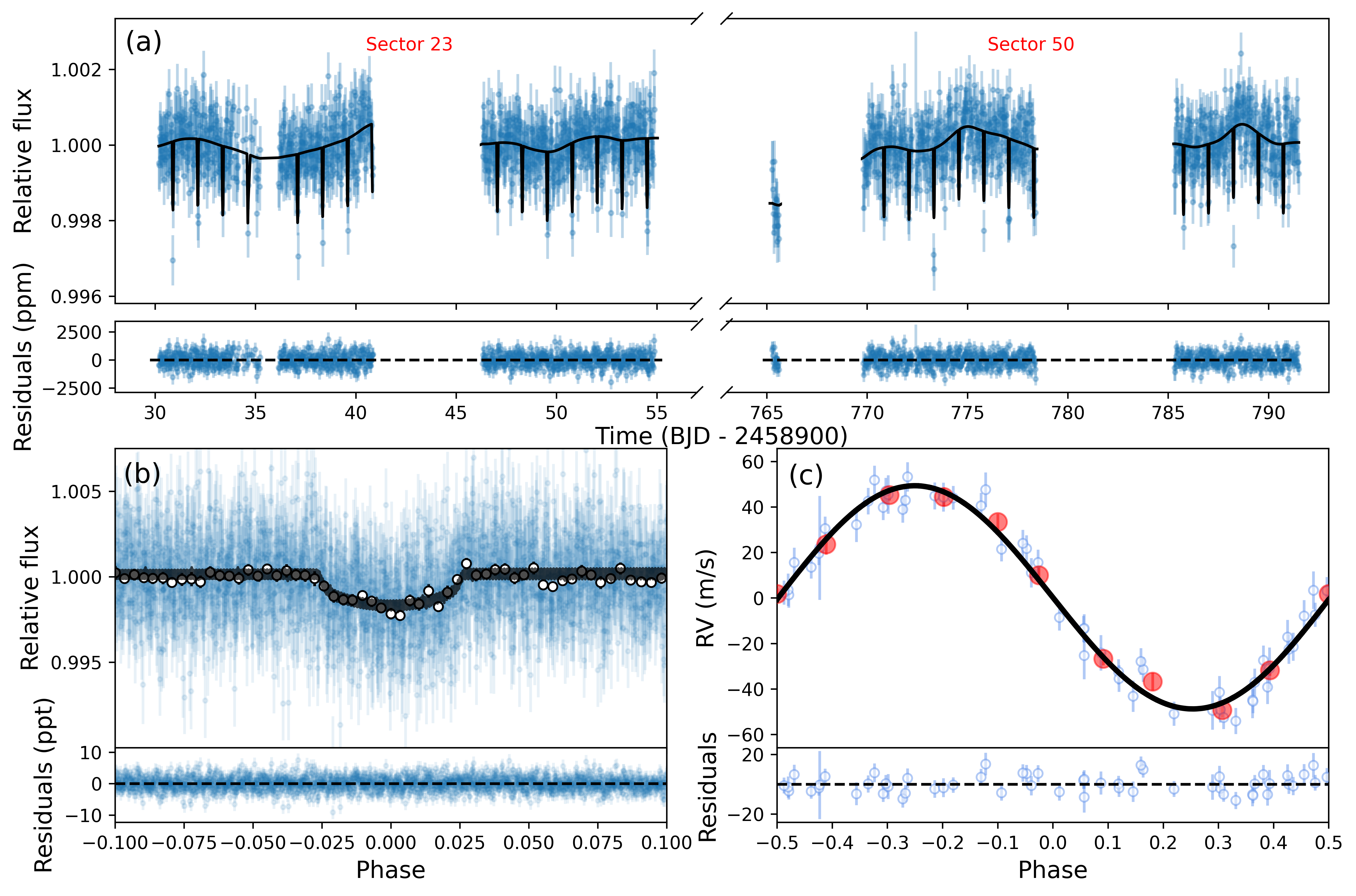}
\end{figure}
\hspace{-0.7cm} \textbf{Figure 1: (a) TOI-1853 light curve, (b) Phase curve of all the transits, (c) Phase curve of radial velocities}. The light curves are binned at 30-minutes cadence for clarity. Cadences marked with anomaly flags (i.e. Coarse Point, Straylight, Impulsive Outliers and Desaturation events) were excluded from the light curve. In panel (c), the average of $\approx 6$ radial velocity measurements are indicated by red dots. In all panels, the error bars represent one standard deviation, while the best-fitting model are shown in black along with its residuals below.

\clearpage

\begin{figure}
\centering
\includegraphics[width=0.9\textwidth]{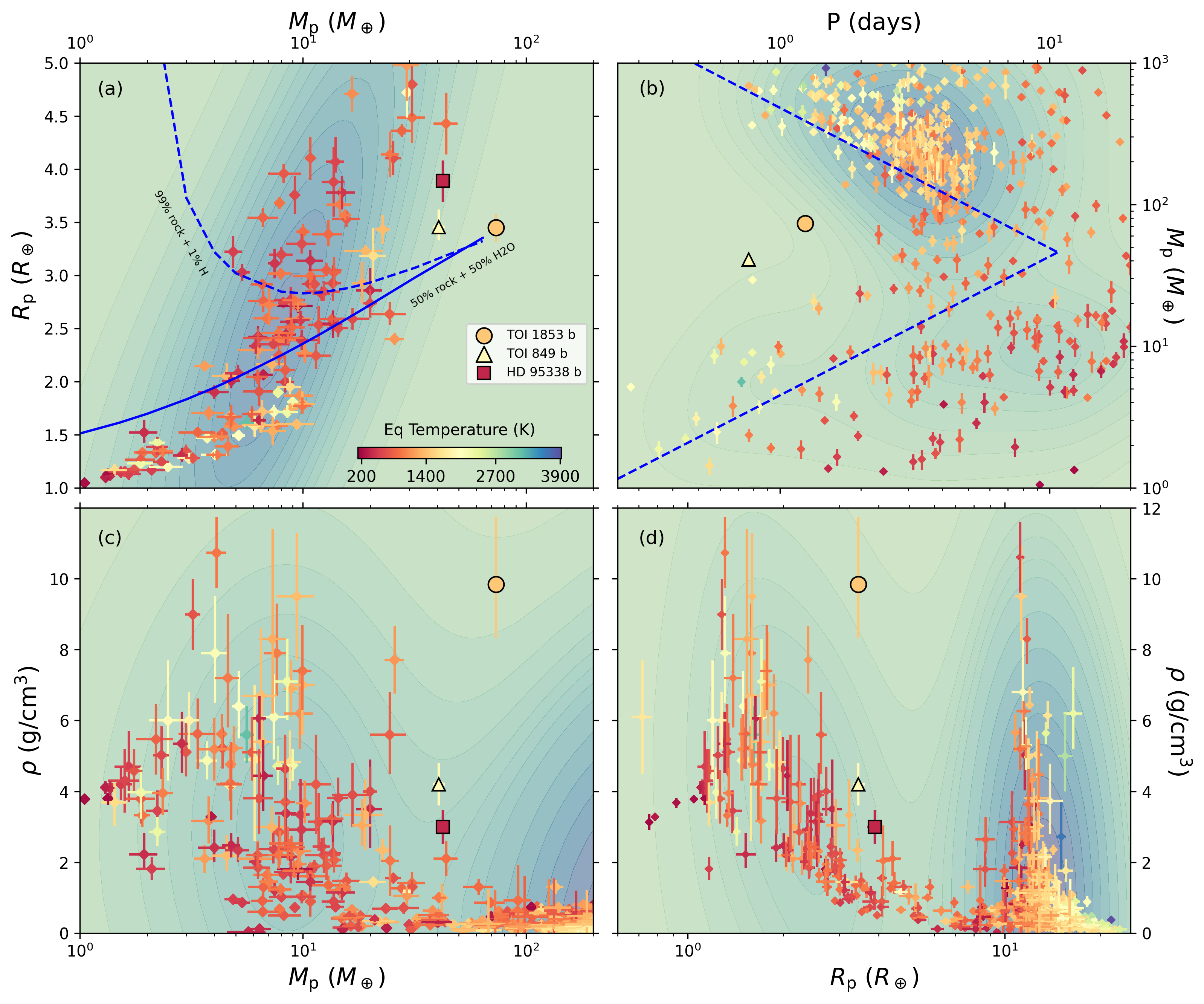}
\end{figure}
\hspace{-0.7cm} \textbf{Figure 2: Diagrams of known transiting exoplanets.} The exoplanets properties have been extracted from TEPCat \cite{Southworth2014} and displayed as diamonds, while their colour is associated with the Equilibrium Temperature. Horizontal and vertical error bars represent one standard deviation. TOI-1853\,b, TOI-849\,b and HD\,95338\,b are displayed as a circle, triangle and square respectively. (a) Radius-mass diagram with blue lines representing different internal compositions [\textbf{dashed}: 99\% Earth-like rocky core + 1\% H layer (at temperature and pressure of 1000K, 1mbar); \textbf{solid}: 50\% Earth-like + 50\% water]. (b) Mass-period diagram, where the blue dashed lines enclose the Neptunian desert \cite{2016A&A...589A..75M}. (c) Density-mass diagram. (d) Density-radius diagram.

\clearpage

\section{Methods}

\vspace{0.5cm}
\subsection{TESS photometric data}
The star TIC\,73540072, or TOI-1853, was observed by \emph{TESS} in Sector 23 at a 30-minute full-frame images (FFI) cadence in early 2020, and in Sector 50 at a 2-minute cadence in early 2022. The transit signal was first identified in the Quick-Look Pipeline (QLP) \cite{huang2020}, and later promoted to \emph{TESS} Object of Interest, TOI-1853.01, planet candidate status by the TESS Science Office \cite{2021ApJS..254...39G}. The Science Processing Operations Center (SPOC) \cite{2016SPIE.9913E..3EJ,Caldwell2020} pipeline retrieved the 2-minutes Simple Aperture Photometry (SAP) and Presearch Data Conditioning Simple Aperture Photometry [PDC-SAP; \cite{Stumpe2014,Smith2012}] light curves. 
The transit signal was identified through the Transiting Planet Search [TPS; \cite{2002ApJ...575..493J,2020ksci.rept....9J}]
and passed all diagnostic tests in the Data Validation [DV; \cite{2018PASP..130f4502T,2019PASP..131b4506L}] modules. All TIC objects, other than the target star, were excluded as sources of the transit signal through the difference image centroid offsets \cite{2018PASP..130f4502T}. Here, for the joint transit-RV analysis of the system, we adopted the PDC-SAP light curve, from which long-term trends were removed using the so-called Co-trending Basis Vectors (CBVs). In particular, for Sector 23 we used the TESS-SPOC High-Level Science Product (HLSP) \cite{Caldwell2020} light curve. 

\subsection{Ground-based photometric follow-up}

The \emph{TESS} pixel scale is $\sim 21$\,arcsec\, pixel$^{-1}$ and photometric apertures extend out to multiple pixels, generally causing multiple stars to blend in the \emph{TESS} aperture. To rule out the NEB-blend scenario 
and attempt to detect the signal on-target, we observed the field as part of the \emph{TESS} Follow-up Observing Program Sub Group 1 [TFOP; \cite{collins:2019}]. 

We observed full transit windows of TOI-1853\,b on May 25, 27 and June 27 2020 respectively with the MuSCAT2 imager \cite{Narita:2019} installed at the 1.52~m Telescopio Carlos Sanchez in the Teide Observatory (in $g$, $r$, $i$, and $z_\mathrm{s}$ bands), with the 0.61\,m University of Louisville Manner Telescope (ULMT) located at Steward Observatory (through a Sloan-$r^{\prime}$ filter) and with the Las Cumbres Observatory Global Telescope [LCOGT; \cite{Brown:2013}] 1.0\,m network node at Siding Spring Observatory (through a Sloan-$g'$ band filter). We extracted the photometric data with {\tt AstroImageJ} \cite{Collins:2017} and 
measured transit-depths across multiple optical bands consistent with an achromatic transit-like event and compatible with \emph{TESS}. We ruled out an NEB blend as the cause of the TOI-1853\,b detection for all the surrounding stars. We additionallyperformed a joint transit model fit using ULMT and LCOGT data (Ext.\,Fig.\,1), including an ephemeris prior from the global fit of this work (Ext.\,Table\,2). We derived $R_{\rm p} = 3.35^{+0.40}_{-0.30}\,R_{\oplus}$, which is well within $1\sigma$ consistent with the value from the global fit.

\subsection{High Resolution Imaging}
As part of the standard process for validating transiting exoplanets to assess the possible contamination of bound or unbound companions on the derived planetary radii \cite{2015ApJ...805...16C}, we observed TOI-1853 with near-infrared adaptive optics (AO) imaging at Keck and with optical speckle imaging at Gemini and SOAR. 

We performed observations at the Keck Observatory with the NIRC2 instrument on Keck-II behind the natural guide star AO system \cite{2000PASP..112..315W} on 2020-05-28 UT in the standard 3-point dither pattern. 
The dither pattern step size was $3^{\prime \prime}$ and was repeated twice, with each dither offset from the previous dither by $0.5^{\prime \prime}$. NIRC2 was used in the narrow-angle mode with a full field of view of $\sim10^{\prime \prime}$ and a pixel scale of approximately $0.0099442''$ per pixel. The Keck observations were made in the narrowband Br-$\gamma$ filter $(\lambda_{\rm O} = 2.1686; \Delta\lambda = 0.0326~\mu$m) with an integration time of 15 seconds for a total of 135 seconds on the target. 
The final resolutions of the combined dithers were determined from the full width at half maximum (FWHM) of the point spread functions: $0.053^{\prime \prime}$. 	
The sensitivities of the final combined AO image (Ext.\,Fig.\,2) were determined by injecting simulated sources azimuthally around the primary target every $20^\circ$ at separations of integer multiples of the central source's FWHM \cite{2017AJ....153...71F}. 

We observed TOI-1835 with the $^{\backprime}$Alopeke speckle imaging camera at Gemini North \cite{2020AJ....159...19Z,2021FrASS...8..138S}, obtaining seven sets of 1000 frames (2020-06-10 UT), with each frame having an integration time of 60~ms, in each of the instrument's two bands (centred at 562~nm and 832~nm) \cite{2011AJ....142...19H}. The observations of the target reach a sensitivity in the blue channel of 5.2~mag, while in the red channel of 6.3~mag at separations of 0.5\,arcsec, and show no evidence of additional point sources. We also searched for stellar companions to TOI-1853 with speckle imaging on the 4.1-m SOAR telescope \cite{2018PASP..130c5002T} on 2021-02-27 UT, observing in Cousins-$I$ band, similar to the TESS bandpass. This observation was sensitive to a 4.7-magnitude fainter star at an angular distance of 1\,arcsec from the target.
	
\subsection{Gaia}
Gaia DR3 astrometry \cite{2021A&A...649A...1G} provides additional information on the possibility of inner companions that may have gone undetected by high-resolution imaging. For TOI-1853 Gaia found a Renormalised Unit Weight Error (RUWE) of $\approx 1$, indicating that the Gaia astrometric solution is consistent with the star being single \cite{2020AJ....159...19Z}. In addition, Gaia identified no widely separated companions that have the same distance and proper motion as TOI-1853.

\subsection{Spectroscopic data}
We gathered 56 spectra of TOI-1853 with HARPS-N [High Accuracy Radial velocity Planet Searcher for the Northern hemisphere; \cite{2012SPIE.8446E..1VC}] between February 2021 and August 2022, within the GAPS programme, and reduced them using the updated online Data Reduction Software (DRS) v2.3.5 \cite{2021plat.confE.106D}.

We observed the star in $\rm OBJ\_AB$ mode, with fiber A on the target and fiber B on the sky to monitor possible contamination by moonlight, which we deemed negligible \cite{2017AJ....153..224M}. We extracted the RVs by cross-correlating the HARPS-N spectra with a stellar template  close to the stellar spectral type. 
The median (mean) of the formal uncertainties of the HARPS-N RVs is 3.8 (4.6) m\,s$^{-1}$; the RV scatter of $34$\,m\,s$^{-1}$ reduces to $4.5$\,m\,s$^{-1}$ after removing the planetary signal.

\subsection{Stellar analysis}

We derived spectroscopic atmospheric parameters exploiting the co-added spectrum of TOI-1853.
In particular, we measured effective temperature ($T_{\rm eff}$), surface gravity ($\log{g}$), microturbulence velocity ($\xi$), and iron abundance ([Fe/H]) through a standard method based on measurements of equivalent widths (EWs) of iron lines [see \cite{biazzoetal2015,biazzoetal2022}]. We then adopted the \cite{castellikurucz2003} grid of model atmospheres with new opacities and the spectral analysis package MOOG \cite{sneden1973} version 2017. $T_{\rm eff}$ was derived by imposing that the abundance of Fe\,\textsc{i} is not dependent on the line excitation potentials, $\xi$ by obtaining the independence between Fe I abundance and the reduced iron line EWs, and $\log{g}$ by the ionization equilibrium condition between Fe\,\textsc{i} and Fe\,\textsc{ii}. 

We also computed the elemental abundance of magnesium and silicon, with respect to the Sun \cite{biazzoetal2022}, using the same code and grid of models. 
The elemental ratios [Mg/Fe] and [Si/Fe] have solar values within the errors, with no evident enrichment in none of these elements with respect to the others. The stellar projected rotational velocity ($v\sin{i} $) was obtained through the spectral-synthesis technique of three regions around 5400, 6200, and 6700\,\AA, as done in \cite{biazzoetal2022}. By assuming a macroturbolence velocity $v_{\rm macro}=1.8$\,km\,s$^{-1}$ from the relationships by \cite{breweretal2016}, we found $v\sin{ i}=1.3\pm0.9$\,km\,s$^{-1}$, which is below the HARPS-N spectral resolution, thus suggesting a slow stellar rotation unless the star is observed nearly pole-on.

Finally, we determined the stellar physical parameters with the EXOFASTv2 tool \cite{2017ascl.soft10003E},
which simultaneously adjusts the stellar radius, mass and age in a Bayesian differential evolution Markov chain Monte Carlo (DE-MCMC) framework \cite{TerBraak2006}, through the modelling of the stellar Spectral Energy Distribution (SED) and the use of the MESA Isochrones and Stellar Tracks (MIST) \cite{2015ApJS..220...15P}. To sample the stellar SED we used the APASS Johnson $B$, $V$ and Sloan $g^{\prime}$, $r^{\prime}$, $i^{\prime}$ magnitudes \cite{2016yCat.2336....0H}, 
the 2-MASS near-infrared $J$, $H$ and $K$ magnitudes \cite{2003tmc..book.....C}, and 
the WISE W1, W2, W3 infrared magnitudes \cite{2014yCat.2328....0C}.
We imposed Gaussian priors on the $T_{\rm eff}$ and [Fe/H] as derived from the analysis of the HARPS-N spectra, 
and on the Gaia DR3 parallax $6.0221\pm0.0159$~mas \cite{2022arXiv220800211G}. 
We used uninformative priors for all the other parameters, including the $V$-band extinction $A_{V}$, for which we adopted an upper limit of 0.085 from reddening maps \cite{2011ApJ...737..103S}. Ext.\,Fig.\,3 displays the best fit of the stellar SED.

\subsection{RV and activity indicators periodograms}
Simultaneously with the RVs, we extracted the time series of several stellar activity indices (Ext.\,Fig.\,4): the FWHM, contrast and bisector (BIS) span of the CCF (Cross-Correlation Function) profile, as well as the Mount Wilson index ($S_{MW}$) and the spectroscopic lines H$\alpha$, Na and Ca. We computed the Generalized Lomb-Scargle (GLS) periodogram, with \texttt{astropy} v.4.3.1 \cite{Zechmeister2009,2018AJ....156..123A}, for both the RVs and the activity indexes, which can be inspected in Ext.\,Fig.\,4. In the RVs, we found the most significant peak at 1.24 days (FAP $\ll$ 0.1\%), i.e. the expected transiting period of TOI-1853\,b. This signal is not attributable to stellar activity since none of the measured activity indicators shows a similar periodicity or harmonics. Another strong peak appears at the 1d alias frequency $f_{\rm alias}$ of the planetary period in the form of $f_{\rm alias}=f_{1\,{\rm d}}-f_{1.24\,{\rm d}}$, giving rise to a period of $P_{\rm alias}\approx5.1$\,days, which is no longer seen when the signal of TOI-1853\,b is subtracted, along with any other peak with $\rm{FAP} \lesssim 5 \%$. The activity indicators do not show signals with $\rm{FAP} \gtrsim 0.1 \%$.

\subsection{Joint transit and RV analysis}
For the joint transit-RV analysis of TOI-1853\,b, we employed the modelling tool \texttt{juliet} \cite{2019MNRAS.490.2262E}, which makes use of \texttt{batman} \cite{2015PASP..127.1161K} for the modelling of transits, \texttt{RadVel} \cite{2018PASP..130d4504F} for the modelling of RVs and correlated variations which are treated as GPs with the packages \texttt{george} \cite{2015ITPAM..38..252A} and \texttt{celerite} \cite{2017AJ....154..220F}. We exploited the dynamic nested sampling package, \texttt{dynesty} \cite{2020MNRAS.493.3132S}, to compute Bayesian posteriors and evidence for the models. 

Even though by default the PDC-SAP photometry is already corrected both for dilution from other objects contained within the aperture using the Compute Optimal Aperture (COA) module \cite{2020ksci.rept....3B}, and major systematic errors, we corrected it for minor fluctuations that were still observable in the light curve (Fig.\,1). In particular, we normalized it by fitting a simple (approximate) Matern GP (Gaussian Process) kernel \cite{2019MNRAS.490.2262E}. We also analysed the SPOC SAP photometry \cite{Twicken2010,MorriS2020}, which is not corrected for long trends and found no significant change in the transit depths, or any conclusive evidence of the stellar rotation period over the brief time coverage of both sectors.

We constructed a transit light curve model with the usual planetary orbital parameters: period $P$, time of inferior conjunction $T_0$, eccentricity $e$ and argument of periastron $\omega$ via the parametrization $(\sqrt{e}\sin{\omega}, \sqrt{e}\cos{\omega})$ 
and the mean density of the parent star $\rho_{\star}$ \cite{2019MNRAS.490.2262E} from our stellar analysis. The \emph{TESS} data parameters adopted in our model were the flux offset and jitter parameters, $TESS_{\rm off}$ and $TESS_{\rm jitt}$, and the two hyper-parameters of the Matern GP model: $\sigma_{\rm GP}$ and $\rho_{\rm GP}$, respectively the amplitude in parts per million (ppm) and the length-scale of the GP. The impact parameter [$b=(a_{\rm p}/R_{\star})\cos{i_{\rm p}}$ for a circular orbit] and the planet-to-star radius ratio $k$ were parameterized as $(r_1,r_2)$ \cite{2018RNAAS...2..209E}. Moreover, here we make use of the limb-darkening parametrizations described in \cite{2013MNRAS.435.2152K} for the two-parameters limb-darkening law ($q_1,q_2 \to u_1,u_2$), with Gaussian priors centred around the values estimated from \cite{Claret2017}. Then, we include the RV model with the usual systemic velocity $\overline{\gamma}_{\rm HARPS-N}$, jitter $\sigma_{\rm HARPS-N}$, and the RV signal semi-amplitude $K_{\rm p}$.

The priors for all the parameters that are used in the joint analysis along with the estimates of the parameters' posteriors are summarized in Ext.\,Table\,2, while the posterior distributions of the sampling parameters are shown as corner plots in Ext.\,Fig.\,5. The addition of linear slopes or RV GP models (quasi-periodic, simple harmonic oscillator, Matern) did not increase the Bayesian-log evidence. We also did not detect any trace of transit time variations in a test done with \texttt{juliet}, where the orbital period of the transit model was fixed at its best-fitting value while all the transit times were allowed to vary. Furthermore, the inclusion of a second planetary RV signal didn't improve the Bayesian-log evidence, which is consistent with the lack of a significant peak in the GLS of the RVs after the subtraction of TOI-1853\,b signal.

\subsection{RV detection function}

We estimated the detection function of the HARPS-N RV time series by performing injection-recovery simulations, in which synthetic planetary signals were injected in the RV residuals after the subtraction of TOI-1853\,b signal. We simulated signals of additional companions in a logarithmic grid of $30 \times 40$ in the planetary mass, $M_{\rm p}$, orbital period, $P$, parameter space respectively, covering the ranges $1-1000 \, M_\oplus$ and $0.5-5000$ d. For each location in the grid, we generated 200 synthetic planetary signals, drawing $P$ and $M_{\rm p}$ from a log-uniform distribution inside the cell, $T_0$ from a uniform distribution in $[0,P]$, and assuming circular orbits. Each synthetic signal was then added to the RV residuals. We computed the detection function as the recovery rate of these signals, i.e. fitting the signals with either a circular-Keplerian orbit or a linear or quadratic trend, to correctly take into account long-period signals which would not be correctly identified as a Keplerian due to the short time-span of the RV observations (500 d). We adopted the Bayesian Information Criterion (BIC)
to compare the fitted planetary model with a constant model: when $\Delta \text{BIC} > 10$ in favour of the planet-induced model, we considered the planetary signal significantly detected. The detection function was then computed as the fraction of detected signals for each element of the grid (Ext.\,Fig.\,6).



\subsection{Orbital decay}
According to the current tidal theory, tidal inertial waves are presently not excited inside TOI-1853 by planet b because the rotation period of the star is certainly longer than twice the planetary orbital period \cite{Ogilvie2007}, while tidal gravity waves are not capable of dissipating efficiently in the central region of the star given the relatively low planetary mass \cite{Barker2020}. Even assuming a lower limit for the modified tidal quality factor $Q'_{\star}=10^7$, the remaining lifetime of the planet is about 4 Gyr according to Eq. (1) of ref. \cite{Metzger2012}, or longer if we take into account equilibrium tides \cite{Collier2018}.

The situation is different if the planet's orbital plane is inclined relative to the stellar equator \cite{Lai2012} because a dynamical obliquity tide is expected to be excited independently of the star rotation period. Such a dynamical tide would produce a fast decay of the obliquity, while it is not equally effective in producing a decay of the orbit semi-major axis \cite{Lai2012}. Therefore, the effective $Q^{\prime}_{*}$ ruling the orbital decay can be assumed to be approximately unaffected by the stellar or planetary obliquities [see, however, Sect. 2.2 of \cite{Leconte2010}, for a quantification of how the obliquities affect the evolution of the semimajor axis]. On the other hand, assuming $Q'_{\star}\sim10^6$ for the dynamical obliquity tide, the e-folding damping time of the stellar obliquity would be about 1.6 Gyr, which is significantly shorter than the estimated age of the system. Therefore, any initial stellar obliquity may have had time to be damped along the lifetime of the system. 

\subsection{Formation simulations}

We simulated systems of 2, 4 and 8 solid-rich planets with a total mass of 80$\,M_{\oplus}$, using the code \texttt{swift\,symba5} \cite{Duncan1998}. For the first scenario of merging proto-planets, we placed the innermost planet at 0.02 au from the star, similar to TOI-1853\,b, and the other planets separated by 1.5 mutual Hill radii, to ensure that the system is violently unstable. The initial eccentricities were assumed to be between 0 and 5e-2 and the inclinations between 0 and 1.4 degrees, to ensure that the system evolves in 3D and the collision probability is not artificially enhanced. The systems with initially 2 and 4 proto-planets merged into a single planet with 80$\,M_{\oplus}$. The system with 8 super-Earths merged into two planets of 50 and 30$\,M_{\oplus}$ respectively. We then did a fourth simulation starting from a system of 10 super-Earths of 10$\,M_{\oplus}$ each, which led to the formation of two planets of 70$\,M_{\oplus}$ and 30$\,M_{\oplus}$.

For the second scenario, we simulated a Jupiter-mass planet on an orbit with a semi-major axis at 1\,au, perihelion distance at 0.02\,au and inclination of 10 degrees. The planet is assumed to have a radius of 2 Jupiter radii, due to tidal heating at perihelion. We placed test particles in three rings at 0.02-0.06\,au, 0.1-0.3\,au and 0.5-1.5\,au. In a million years, the planet engulfed 30\%, 6\% and 2\% of the particles in the three rings, respectively. The sharp decay of efficiency with the distance from the star is expected from the increase in Savronov's number \cite{Safronov1972}.

\subsection{Detailed impact simulations}

A sequence of giant impacts should remove most of the gaseous envelopes that might have been initially present around the original proto-planets. We conducted a series of smoothed particle hydrodynamics impact simulations to assess the effectiveness of atmosphere removal under different conditions, using \texttt{SWIFT} \cite{schaller_swift_2016,kegerreis_planetary_2019}. Assuming the planets have three layers (H/He on top, rock in the core and water in between) we explored three different compositions: water-rich by mass (67.5\,\% water, 22.5\,\% rock, 10\,\% H/He), neutral (45\,\% rock, 45\,\% water, 10\,\% H/He), and rock-rich (67.5\,\% rock, 22.5\,\% water, 10\,\% H/He), with the planets' thermodynamic profiles generated using \texttt{WoMa} \cite{kegerreis_planetary_2019}. The rocky core, water, and atmosphere layers are modelled using the ANEOS forsterite \cite{stewart_shock_2020}, AQUA water \cite{haldemann_aqua_2020}, and a mixture of hydrogen–helium \cite{hubbard_structure_1980} equation of states, respectively. The ANEOS forsterite equation of state table was regenerated using \cite{stewart_equation_2019}.
The mass of the largest post-collision remnant ($M_{\rm lr}$) was calculated using the method described in \cite{marcus_collisional_2009,carter_collisional_2018}. The maximum smoothing length $h_{\rm max}$ was set to $5\, R_{\oplus}$ in these \texttt{SWIFT} simulations.

We tested various impact scenarios including head-on and oblique impacts with pre-impact impactor to target mass ratio ranging from 0.5 to 1 (target masses are 25, 45, 50 and 60 $M_{\oplus}$). We found that head-on (oblique) merging collisions, where the target and impactor collide at a speed approximately equal to their mutual escape velocity, would result in the removal of at most 10\% (5\%) of their atmospheres. Therefore, the system would have to go through at least one high-speed impact ($\gtrsim2$ mutual escape velocity) during its final formation stage in order to remove most of the atmosphere. In Ext.\,Table\,3, we list three impact simulation results that could lead to the formation of TOI-1853\,b.

A high percentage of water on the initial planets would make the removal of the atmosphere more efficient, as water-rich impacts require the lowest impact velocity and have the highest atmosphere loss fraction. Although there is always some loss of water during impacts, the loss efficiency is reduced if more water is present on the planet initially. 
The mass fractions of the rock in the post-collision remnant tend to increase in all three impact setups, confirming that rock is less likely to be lost due to the core being in the centre and being less compressible than water and the atmosphere.

\subsection{Photo-evaporation}
At the present age, the Jeans escape parameter $\Lambda$ ($\propto M_{\rm p} / R_{\rm p} T_{\rm eq}$) \cite{Fossati+2017} indicates that the atmosphere of TOI-1853\,b is in hydrodynamic stability, thanks to the deep gravitational potential well of the planet. 
To explore its evolution, we adapted a numerical code developed for studying single systems \cite{Locci+2019,Maggio+2022}, considering the following three scenarios: rocky core + an H$_2$-dominated envelope (1\% by mass), 49.95\% rocky core + 49.95\% water mantle + 0.1\% H$_2$-dominated envelope, 50\% rocky core + 50\% water mantle and no envelope. For each case, we created a synthetic population of young planets (10\,Myr old) with a large envelope and different atmospheric fractions. At any age, the planet's radius is the sum of a fixed core radius plus a time-dependent envelope radius.

The planets passed through an evaporation phase in which the radius contracted due to mass loss and gravitational shrinking. We followed their evolutionary history up to 7\,Gyr and found that none of the planets is either initially, or later becomes, hydro-dynamically unstable. Larger radii imply lower values of the Jeans escape parameter, but this is compensated by the lower equilibrium temperatures until the host star reaches the main sequence at an age of $\sim 200$\,Myr. This means that the current planet's structure cannot be the result of strong photo-evaporation processes.

\subsection{Atmospheric transmission spectrum modeling}

To estimate expected transmission spectral signals that could be observed by the JWST \cite{JWST}, we generated synthetic spectra for TOI-1853\,b using the open-source \textsc{Pyrat Bay} modelling framework \cite{CubillosBlecic2021mnrasPyratBay}.
We considered two extreme cases based on the scenarios previously explored: 99\% rocky core with 1\% H$_2$ atmosphere, and 50\% rocky core, 40\% H$_2$O mantle with a 10\% H$_2$O supercritical steam atmosphere.
We assumed an atmosphere with an isothermal temperature profile at the planet's $T_{\rm eq}$ and, for simplicity, we only considered the opacity from H$_2$ and H$_2$O. We then simulated with \texttt{PandExo} \cite{BatalhaEtal2017paspPandexo} a series of transmission-spectrum data for the JWST NIRISS/SOSS, NIRSpec/G395H, and MIRI/LRS to cover the $0.9-10.0\,\mu$m wavelength range.

Ext.\,Fig.\,7 shows our model spectra of TOI-1853\,b along with the simulated JWST observations combining three transits with each instrument to enhance the signal-to-noise ratios. The main difference between the H$_2$- and H$_2$O-dominated cases is the flattening of the H$_2$O spectral features due to the larger mean molecular mass of the latter. For the H$_2$-dominated atmosphere we assumed a solar H$_2$O trace abundance, though there were no significant variations for a range $0.01-50.0\times$\,solar metallicities. Of the three instruments, NIRISS/SOSS is the best suited to differentiate between H$_2$- and H$_2$O-dominated atmospheres, being able to trace the slope of increasing H$_2$O absorption with wavelength over the 0.9 and 2.8 $\mu$m range for the H$_2$-dominated atmosphere, as opposed to the flat transmission spectrum for the H$_2$O-dominated atmosphere.


\bibliographystyle{Science}
\bibliography{bibMT}

\section{Addendum}
\subsection{Author contributions}
L.\,N. analyzed the transit, RV data and wrote the manuscript. A.\,S. and A.\,S.\,B. performed the selection of TESS Neptunes for the HARPS-N follow-up and scheduled the HARPS-N observations within the GAPS consortium. A.\,S. performed a preliminary RV analysis. L.\,M., A.\,S.\,B., A.\,S. and M.\,D. supervised the work and contributed to writing the manuscript. X.\,D. reduced HARPS-N spectra. M.\,P. estimated the detection function of HARPS-N RVs. A.\,S.\,B. and K.\,B. determined the stellar parameters. A.\,W.\,M. and C.\,Z. performed and analyzed SOAR observations, while J.\,E.\,S., S.\,B.\,H., K.\,V.\,L. and R.\,M. obtained and reduced the Gemini data. D.\,R.\,C and C.\,Z. contributed to writing the high-resolution imaging section. E.\,G. is the Keck data collector. A.\,M. performed the simulations and contributed to writing the formation scenario with the help of J.\,J.\,L., while J.\,D., Z.\,L. and P.\,C. computed the body collision simulations. L.\,Z. analyzed the planet's composition. K.\,A.\,C. scheduled the LCO observations, performed data reduction along with R.\,P.\,S., and contributed to writing the light curve follow-up sections. J.\,F.\,K. performed the ULMT observations and their data reduction. E.\,L.\,N.\,J. performed the joint MCMC analysis of the LCO and ULMT light curves. E.\,Pal. obtained the data of MuSCAT2. D.\,L. and A.\,Ma. analyzed the evolutionary history of the atmosphere. P.\,C. estimated the transmission spectral signals observed by JWST. A.\,L. computed the lifetime of the planet. D.\,S., A.\,B., D.\,N., I.\,P., L.\,Mal. and T.\,Z. are members of the Science Team of the GAPS (Global Architecture of Planetary System) consortium, while A.\,G., R.\,C., W.\,B., A.\,F.\,M.\,F., M.\,Pe. and A.\,H. are members of the TNG (Telescopio Nazionale Galileo) which has conducted the RV observations. L.\,G.\,B. is a member of the \emph{TESS} Payload Operations Center (POC), while J.\,D.\,T. is a member of the \emph{TESS} Science Processing Operations Center (SPOC). A.\,Sh., M.\,B.\,L, S.\,S. and J.\,N.\,W. are \emph{TESS} contributors. All authors have contributed to the interpretation of the data and the results. 

\subsection{Acknowledgments}
We acknowledge the use of public TESS data from pipelines at the TESS Science Office and at the TESS Science Processing Operations Center. Resources supporting this work were provided by the NASA High-End Computing (HEC) Program through the NASA Advanced Supercomputing (NAS) Division at Ames Research Center for the production of the SPOC data products. The work is based on observations made with the Italian Telescopio Nazionale Galileo (TNG) operated on the island of La Palma by the Fundacion Galileo Galilei of the INAF (Istituto Nazionale di Astrofisica) at the Spanish Observatorio del Roque de los Muchachos of the Instituto de Astrofisica de Canarias. This work has also made use of data from the European Space Agency (ESA) mission {\it Gaia} (\url{https://www.cosmos.esa.int/gaia}), processed by the {\it Gaia} Data Processing and Analysis Consortium (DPAC, \url{https://www.cosmos.esa.int/web/gaia/dpac/consortium}). This work makes use of observations from the LCOGT network. Part of the LCOGT telescope time was granted by NOIRLab through the Mid-Scale Innovations Program (MSIP). MSIP is funded by NSF. This research has made use of the Exoplanet Follow-up Observation Program (ExoFOP; DOI: 10.26134/ExoFOP5) website, which is operated by the California Institute of Technology, under contract with the National Aeronautics and Space Administration under the Exoplanet Exploration Program. This paper makes use of observations made with the MuSCAT2 instrument, developed by the Astrobiology Center, at Telescopio Carlos Sánchez operated on the island of Tenerife by the IAC in the Spanish Observatorio del Teide, and is also based in part on observations obtained at the Southern Astrophysical Research (SOAR) telescope, which is a joint project of the Ministério da Ciência, Tecnologia e Inovações do Brasil (MCTI/LNA), the US National Science Foundation’s NOIRLab, the University of North Carolina at Chapel Hill (UNC), and Michigan State University (MSU). This work has been carried out within the framework of the NCCR PlanetS supported by the Swiss National Science Foundation under grants 51NF40-182901 and 51NF40-205606. This paper made use of observation from the High-Resolution Imaging instrument ‘Alopeke and were obtained under Gemini LLP Proposal Number: GN/S-2021A-LP-105. ‘Alopeke was funded by the NASA Exoplanet Exploration Program and built at the NASA Ames Research Center by Steve B. Howell, Nic Scott, Elliott P. Horch, and Emmett Quigley. ‘Alopeke was mounted on the Gemini North telescope of the international Gemini Observatory, a program of NSF’s OIR Lab, which is managed by the Association of Universities for Research in Astronomy (AURA) under a cooperative agreement with the National Science Foundation. On behalf of the Gemini partnership: the National Science Foundation (United States), National Research Council (Canada), Agencia Nacional de Investigación y Desarrollo (Chile), Ministerio de Ciencia, Tecnología e Innovación (Argentina), Ministério da Ciência, Tecnologia, Inovações e Comunicações (Brazil), and Korea Astronomy and Space Science Institute (Republic of Korea). The giant impact simulations were carried out using the computational facilities of the Advanced Computing Research Centre, University of Bristol.

\subsection{Funding}
Funding for the DPAC has been provided by national institutions, in particular, the institutions participating in the {\it Gaia} Multilateral Agreement. L.\,N. and D.\,L. acknowledge the support of the ARIEL ASI-INAF agreement 2021-5-HH.0. L.\,M. acknowledges support from the ``Fondi di Ricerca Scientifica d'Ateneo 2021'' of the University of Rome ``Tor Vergata''. A.\,Ma. and A.\,S.\,B. acknowledge support from the ASI-INAF agreement n.2018-16-HH.0 (THE StellaR PAth project), and from PRIN INAF 2019. Funding for the TESS mission is provided by NASA's Science Mission Directorate. This project has received funding from the European Research Council (ERC) under the European Union’s Horizon 2020 research and innovation programme (grant agreement SCORE No 851555). S.\,B.\,H acknowledges funding from the NASA Exoplanet Program Office. K.\,A.\,C acknowledges support from the TESS mission via sub-award s3449 from MIT. TZi acknowledges support from CHEOPS ASI-INAF agreement n. 2019-29-HH.0. J.\,D. acknowledges the funding support from the Chinese Scholarship Council (No. 202008610218,). This work is partly supported by JSPS KAKENHI Grant Numbers JP17H04574, JP18H05439, and JST CREST Grant Number JPMJCR1761.

\subsection{Competing interests:}
We declare no competing interests.

\subsection{Data and materials availability}
The \texttt{juliet} code is freely available at https://github.com/nespinoza/juliet. \texttt{astropy} is a common core package for astronomy in Python, and \texttt{EXOFASTV2} is a public exoplanet fitting software. TESS photometric time series can be freely obtained from the Mikulski Archive for Space Telescopes (MAST) archive at https://exo.mast.stsci.edu/. All follow-up light curve data are available on the {\tt EXOFOP-TESS} website. Radial velocities are presented in Ext.\,Table\,1. 

\section{Extended data}
\begin{table}
\caption{\textbf{Extended Data Table\,1: HARPS-N Radial velocities}. FWHM is the Full Width Half Maximum of the CCF profile, BIS is the Bisector Inverse Slope span, Contrast is referred to stellar line measurement of the spectral lines and $S_{MW}$ is the Mount Wilson index.}
\centering
\begin{tabular}{ccrcrccc} \\
\hline \\ [-10pt]
BJD$_{\rm TDB}$ & RV & $\sigma_{\rm RV}~~~$ & FWHM & BIS & Contrast & $S_{\rm MW}$ & $\sigma_{S_{\rm MW}}$ \\
(2\,459\,000 + ) & (m\,s$^{-1}$) & (m\,s$^{-1}$) &  &  &  &  &    \\
\hline \\ [-8pt]
277.591475 & -26\,221.79 &  4.97~~ & 6696 & -24 & 66.70484 & 0.3101 & 0.0059 \\
286.743959 & -26\,312.69 &  7.21~~ & 6658 & -37 & 67.19193 & 0.2436 & 0.0096 \\
288.667942 & -26\,223.50 &  4.32~~ & 6656 & -37 & 66.92455 & 0.2719 & 0.0048 \\
289.698557 & -26\,266.22 &  4.32~~ & 6646 & -53 & 66.86836 & 0.2740 & 0.0046 \\
297.604184 & -26\,219.81 &  8.57~~ & 6795 & -7  & 65.98741 & 0.1466 & 0.0102 \\
298.627811 & -26\,222.97 & 10.08~~ & 6720 & -28 & 66.33820 & 0.4311 & 0.0126 \\
299.645873 & -26\,264.59 &  7.65~~ & 6748 & -42 & 66.35285 & 0.3371 & 0.0086 \\
305.749103 & -26\,286.43 &  6.23~~ & 6697 & -22 & 66.60946 & 0.3230 & 0.0078 \\
307.671216 & -26\,246.47 &  4.47~~ & 6743 & -39 & 66.21211 & 0.3493 & 0.0041 \\
325.603943 & -26\,309.55 & 11.53~~ & 6667 & -68 & 66.49581 & 0.3380 & 0.0158 \\
327.671492 & -26\,281.97 &  4.65~~ & 6757 & -32 & 66.08518 & 0.2944 & 0.0045 \\
360.503076 & -26\,287.33 &  7.28~~ & 6789 & -10 & 65.92973 & 0.3035 & 0.0072 \\
365.572246 & -26\,253.24 &  6.43~~ & 6723 & -37 & 66.24472 & 0.3000 & 0.0075 \\
366.609855 & -26\,306.84 &  5.35~~ & 6734 & -47 & 66.30966 & 0.2581 & 0.0064 \\
367.546558 & -26\,299.87 &  4.04~~ & 6695 & -34 & 66.79568 & 0.2893 & 0.0046 \\
377.551116 & -26\,298.74 &  3.69~~ & 6712 & -39 & 66.35716 & 0.3009 & 0.0037 \\
378.532182 & -26\,245.68 &  4.70~~ & 6735 & -41 & 66.01514 & 0.2932 & 0.0047 \\
379.507227 & -26\,208.93 &  6.26~~ & 6748 & -29 & 66.11978 & 0.3012 & 0.0070 \\
387.495223 & -26\,304.60 &  4.86~~ & 6705 & -36 & 66.52209 & 0.2931 & 0.0058 \\
391.502386 & -26\,298.61 &  6.35~~ & 6721 & -24 & 66.32581 & 0.3742 & 0.0076 \\
416.441714 & -26\,287.90 &  5.64~~ & 6683 & -37 & 66.62051 & 0.2863 & 0.0069 \\
565.753307 & -26\,266.64 &  5.33~~ & 6711 & -46 & 66.64815 & 0.3154 & 0.0082 \\
615.721992 & -26\,217.35 &  5.88~~ & 6621 & -38 & 67.06423 & 0.2632 & 0.0093 \\
617.727567 & -26\,312.22 & 10.41~~ & 6700 & -65 & 66.12045 & 0.2601 & 0.0156 \\
618.719631 & -26\,293.86 &  8.97~~ & 6703 & -48 & 66.26545 & 0.2636 & 0.0138 \\
624.713851 & -26\,249.18 &  3.55~~ & 6638 & -39 & 67.17890 & 0.2693 & 0.0050 \\
625.734802 & -26\,232.57 &  4.89~~ & 6642 & -48 & 66.95316 & 0.3119 & 0.0075 \\
627.729205 & -26\,322.66 &  4.68~~ & 6675 & -44 & 66.86213 & 0.3163 & 0.0069 \\
642.724298 & -26\,301.89 &  8.37~~ & 6692 & -61 & 66.50986 & 0.3280 & 0.0139 \\
646.687152 & -26\,249.38 &  2.91~~ & 6697 & -45 & 66.83894 & 0.3069 & 0.0035 \\
\hline
\end{tabular}
\end{table}
\clearpage
\textbf{Extended Data Table\,1.} Continued
\begin{table}
\centering
\begin{tabular}{ccrcrccc}
\hline \\ [-8pt]
648.731285 & -26\,318.04 &  3.59~~  & 6682 & -47 & 66.86892 & 0.2862 & 0.0049 \\
649.716835 & -26\,272.73 &  4.75~~  & 6695 & -41 & 66.83616 & 0.2651 & 0.0070 \\
650.721339 & -26\,223.27 &  3.14~~  & 6665 & -35 & 67.02990 & 0.3126 & 0.0042 \\
656.721298 & -26\,240.13 &  8.62~~  & 6685 & -43 & 66.52727 & 0.2212 & 0.0139 \\
663.759353 & -26\,313.20 &  4.18~~  & 6697 & -53 & 66.54938 & 0.3136 & 0.0057 \\
678.681216 & -26\,304.51 &  7.85~~  & 6664 & -14 & 66.59712 & 0.2085 & 0.0131 \\
681.703784 & -26\,220.95 &  4.43~~  & 6710 & -43 & 66.56025 & 0.3083 & 0.0063 \\
682.627666 & -26\,272.22 &  3.47~~  & 6673 & -31 & 66.80372 & 0.3234 & 0.0047 \\
690.693058 & -26\,253.74 &  6.26~~  & 6734 & -40 & 66.21964 & 0.3055 & 0.0085 \\
705.597822 & -26\,243.73 &  5.00~~  & 6706 & -38 & 66.47439 & 0.3220 & 0.0069 \\
706.532481 & -26\,222.90 &  4.93~~  & 6691 & -41 & 66.80752 & 0.3202 & 0.0074 \\
708.602432 & -26\,321.66 &  9.21~~  & 6705 & -84 & 66.49706 & 0.3678 & 0.0151 \\
717.556786 & -26\,253.74 &  2.63~~  & 6692 & -37 & 66.83179 & 0.3263 & 0.0029 \\
718.552650 & -26\,312.70 &  3.47~~  & 6707 & -46 & 66.61027 & 0.3156 & 0.0041 \\
735.582824 & -26\,291.38 & 13.70~~  & 6701 & -81 & 66.50634 & 0.3421 & 0.0226 \\
737.473873 & -26\,272.62 & 34.93~~  & 6692 &  24 & 65.65718 & 0.0604 & 0.0619 \\
738.528763 & -26\,290.10 &  8.60~~  & 6694 & -37 & 66.51006 & 0.3167 & 0.0134 \\
748.535890 & -26\,270.21 &  9.92~~  & 6700 & -51 & 66.38869 & 0.2286 & 0.0155 \\
750.506612 & -26\,279.02 &  5.85~~  & 6717 & -33 & 66.28337 & 0.2809 & 0.0075 \\
751.517693 & -26\,228.38 &  4.37~~  & 6731 & -39 & 66.26938 & 0.2608 & 0.0052 \\ 
752.506892 & -26\,228.51 &  4.88~~  & 6685 & -40 & 66.58633 & 0.2531 & 0.0070 \\
773.458642 & -26\,265.56 &  3.49~~  & 6693 & -48 & 66.6854 & 0.3202 & 0.0042 \\
774.449876 & -26\,320.37 &  2.72~~  & 6662 & -56 & 66.99418 & 0.3100 & 0.0033 \\
775.457294 & -26\,295.77 &  4.99~~  & 6689 & -42 & 66.72325 & 0.3196 & 0.0072 \\
803.398361 & -26\,245.88 &  3.47~~  & 6656 & -32 & 66.94536 & 0.2849 & 0.0047 \\
807.396243 & -26\,222.41 &  6.10~~  & 6705 & -38 & 66.42269 & 0.2925 & 0.0095 \\
\hline
\end{tabular}
\end{table}
\clearpage

\begin{table}
\caption{\textbf{Extended Data Table\,2: Global joint fit priors and posteriors}. The best-fitting values and uncertainties are extracted from the posterior probability distributions. $\mathcal{U}(a,\:b)$ indicates a uniform distribution between $a$ and $b$; $\mathcal{L}(a,\:b)$ a log-normal distribution, and $\mathcal{N}(a,\:b)$ a normal distribution with mean $a$ and standard deviation~$b$.}
\centering
\begin{tabular}{lccc}
\\
 & & \textbf{Prior} & \textbf{Posterior} \\
\textbf{Parameters} & \textbf{Unit} &   \textbf{distribution} & \textbf{Value ($\pm 1\sigma$)} \\
\hline \\ [-8pt]
{\it Keplerian Parameters} & &  \\
$P$                    & days             & $\mathcal{U}(1,\:100)$ & $1.2436275^{+0.0000027}_{-0.0000031}$               \\
$T_{0}$                & BJD$_{\rm TDB}$  & $\mathcal{U}(2459669.0,\:2459691.5)$ & $2\,459\,690.7419\pm0.0008$ \\
$\sqrt{e}\,\sin\omega$ &                  & $\mathcal{U}(-1.0,\:1.0)$ & $-0.038\pm0.11$            \\
$\sqrt{e}\,\cos\omega$ &                  & $\mathcal{U}(-1.0,\:1.0)$ & $0.043\pm0.06$            \\
$\rho_{\star}$         & kg\,m$^{-3}$     & $\mathcal{N}(2240,\:130)$ & $2255^{+131}_{-128}$ \\ [6pt]
{\it Transit Parameters} & &  \\
$r_1$ &  & $\mathcal{U}(0.0,\:1.0)$ & $0.670\pm^{+0.038}_{-0.045}$   \\
$r_2$ &  & $\mathcal{U}(0.0,\:1.0)$ & $0.0391\pm^{+0.0010}_{-0.0011}$   \\
$q_1$ &  & $\mathcal{N}(0.44,\:0.10)$ & $0.48^{+0.10}_{-0.09}$ \\
$q_2$ &  & $\mathcal{N}(0.20,\:0.10)$ & $0.25^{+0.09}_{-0.10}$ \\
$b$ &     & -- & $0.51^{+0.06}_{-0.07}$ \\ 
$i$ & deg & -- & $84.9^{+0.7}_{-0.6}$ \\
$TESS_{\textsf{off}}$  & & $\mathcal{N}(1.0,\:0.001)$ & $0.0001\pm0.0002$ \\
$TESS_{\textsf{jitt}}$ & ppm  & $\mathcal{L}(10^{-1},\:10^3)$ & $0.27^{+2.77}_{-0.24}$ \\
$\sigma_{\textsf{GP}}$ & ppm  & $\mathcal{L}(10^{-5},\:10)$ & $0.00052^{+0.00016}_{-0.00010}$   \\
$\rho_{\textsf{GP}}$   & days & $\mathcal{L}(0.1,\:100)$ & $3.0^{+1.0}_{-0.7}$ \\  [6pt]
{\it RV Parameters} & &  \\
$K_{\rm p}$                         & m\,s$^{-1}$ & $\mathcal{U}(0.0,\:100.0)$ & $49.0\pm1.2$     \\
$\overline{\gamma}_{\textsf{HARPS-N}}$ & m\,s$^{-1}$ & $\mathcal{U}(-26280,\:-26250)$ & $-26267.4\pm0.8$ \\
$\sigma_{\textsf{HARPS-N}}$         & m\,s$^{-1}$ & $\mathcal{U}(0,\:10)$ & $4.5^{+0.8}_{-0.7}$          \\
\hline
\end{tabular}
\end{table}

\clearpage

\begin{table}
\caption{\textbf{Extended Data Table\,3: Impact simulation results.} Three scenarios of a $30\,M_{\oplus}$ impactor colliding onto a $60\,M_{\oplus}$ target with varying water/rock ratios. $M_{\rm lr}$ is the mass of the largest post-collision remnant. $f$ shows the mass ratio of each material in the largest post-collision remnant. $X^{\rm atmos}_{\rm loss}$ shows the total mass fraction of the atmosphere that is lost. All simulations shown are head-on impacts. }
\centering
\label{tab:impact_data}
\begin{tabular}{ccccccc}
\\
\hline \\ [-10pt]
Component & $V_{\rm imp}$\,(km/s) & $M_{\rm lr}/M_{\oplus}$ & $f_{\rm H_{2}O}$ & $f_{\rm si}$ & $f_{\rm H\&He}$ & $X^{\rm atmos}_{\rm loss}$ \\
\hline \\ [-8pt]
Water Rich           & 75.5 & 74.2  & 71.67\% & 27.20\% & 1.13\% & 90.7\% \\
Equal water and rock & 77.7  & 72.8  & 42.92\% & 55.30\% & 1.78\% & 85.6\% \\
Rock Rich            & 80.2 & 73.5  & 16.13\% & 81.11\% & 2.76\% & 77.4\% \\
\hline
\end{tabular}
\end{table}

\clearpage



\begin{figure}
\centering
\includegraphics[width=1\columnwidth]{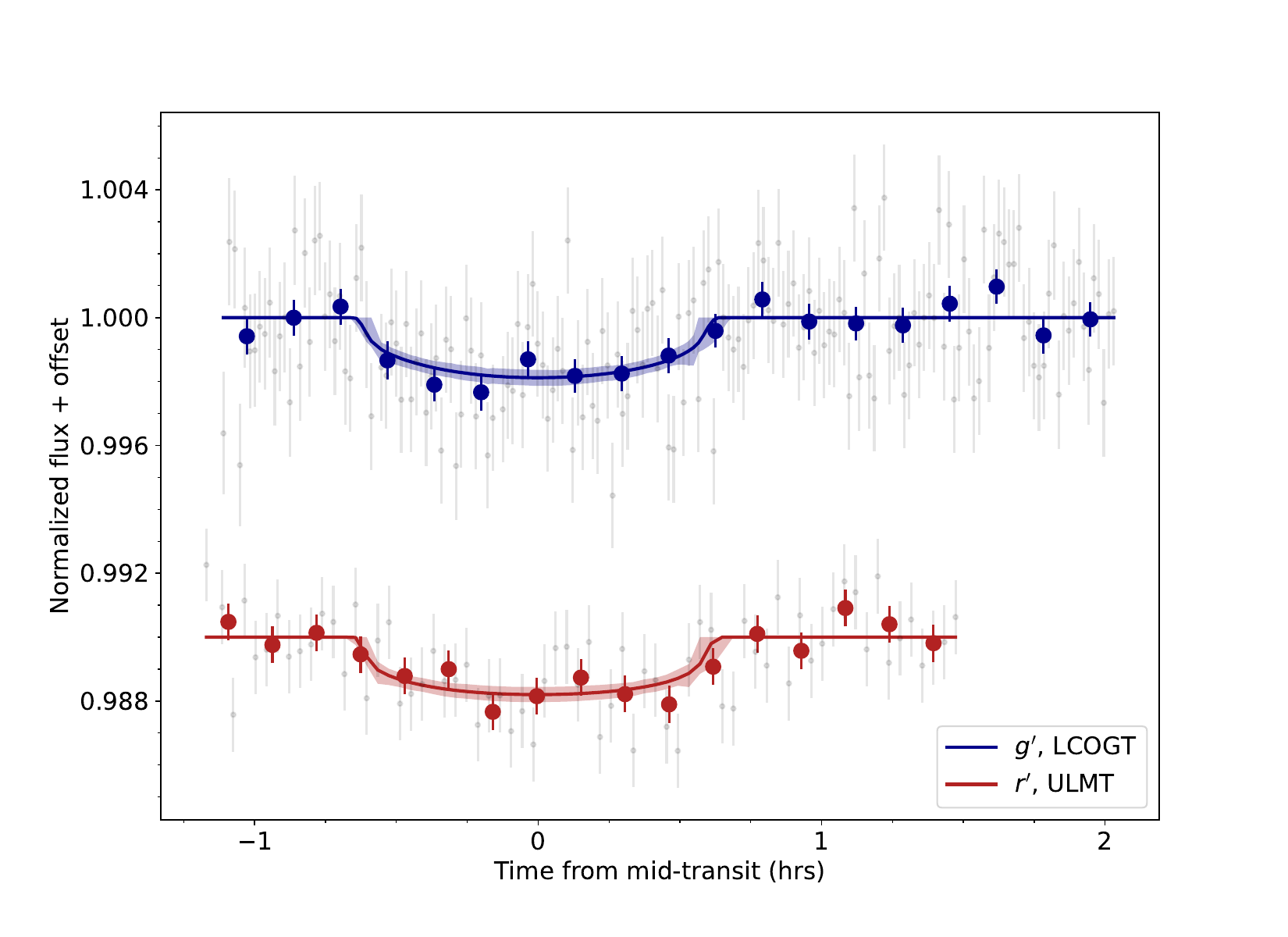}
\end{figure}
\hspace{-0.7cm} \textbf{Extended Data Figure 1: Fit of phase-folded on-ground transits}. The blue symbols show the LCOGT data in Sloan $g'$ band in 10-minute bins. The red symbols show the ULMT data in Sloan $r'$ band in 10-minute bins, shifted on the Y-axis for clarity. Grey symbols show the unbinned data. The error bars represent one standard deviation. A joint transit model fit to both light curves, which included an ephemeris prior from the global fit in this work, is over-plotted using solid lines. 
\clearpage

\begin{figure}
\centering
\includegraphics[width=1.0\textwidth]{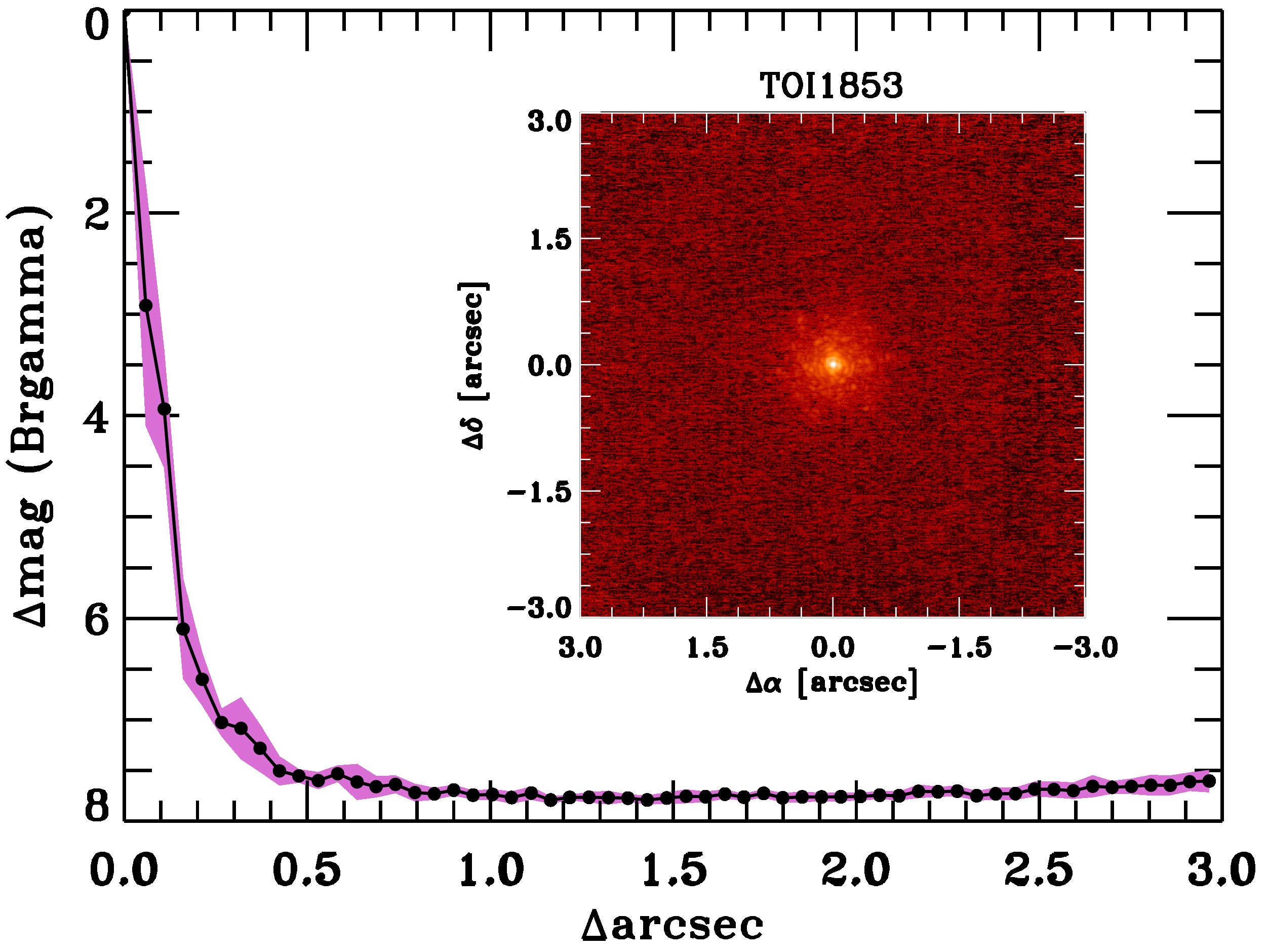}
\end{figure}
\hspace{-0.7cm} \textbf{Extended Data Figure 2: Keck Near-Infrared Adaptive Optics sensitivity curve.} The image reaches a contrast of $\sim 7.5$ magnitudes fainter than the host star within $0.5^{\prime \prime}$ in the Br-$\gamma$ filter. {\it Inset:} Image of the central portion of the data.
\clearpage

\begin{figure}
\centering
\includegraphics[width=16cm]{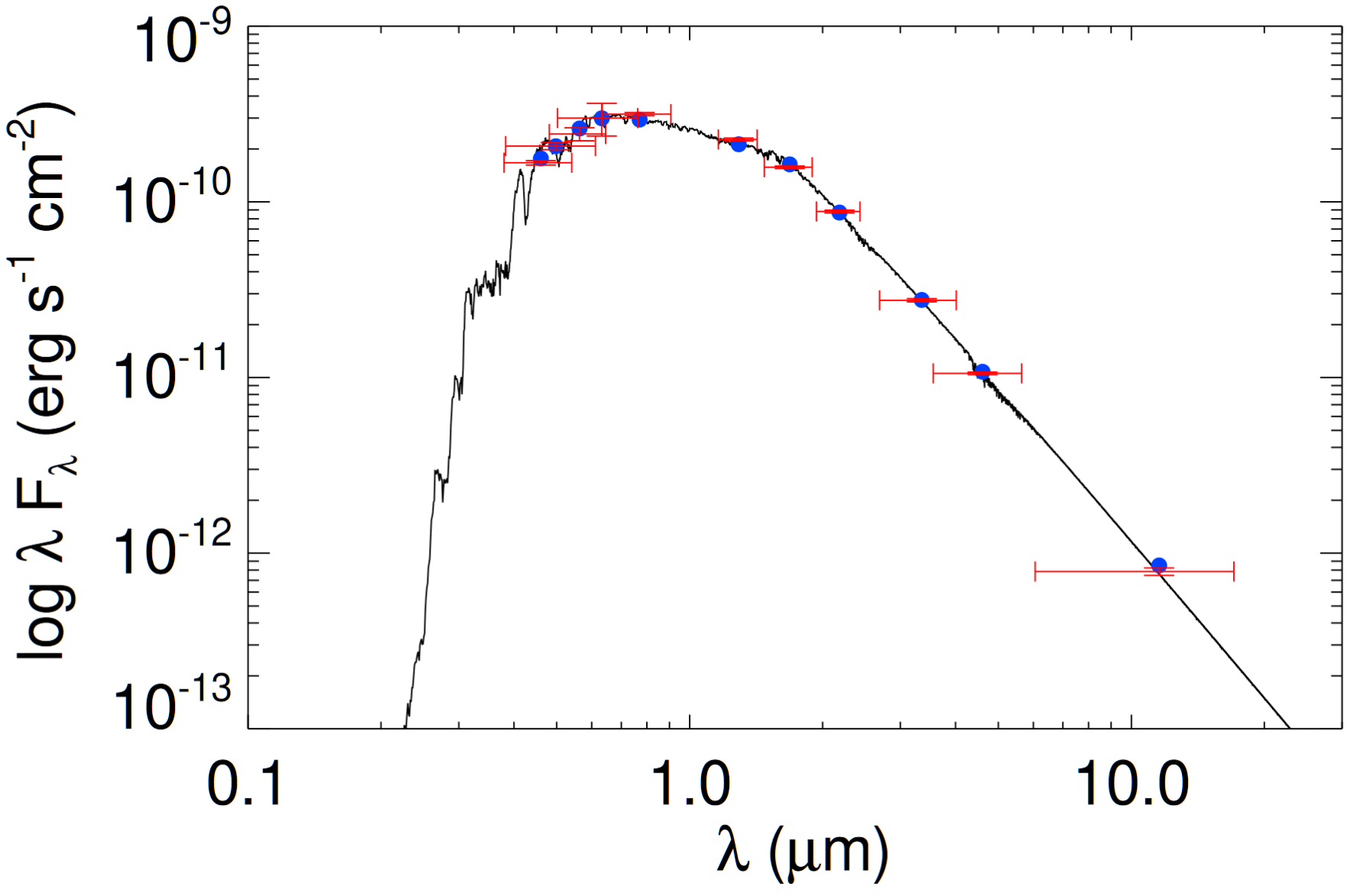}
\end{figure}
\hspace{-0.7cm} \textbf{Extended Data Figure 3: TOI-1853 Spectral Energy Distribution}. The error bars represent one standard deviation. The best-fit model is displayed as a solid black line.
\clearpage

\begin{figure}
\centering
\includegraphics[width=0.75\textwidth]{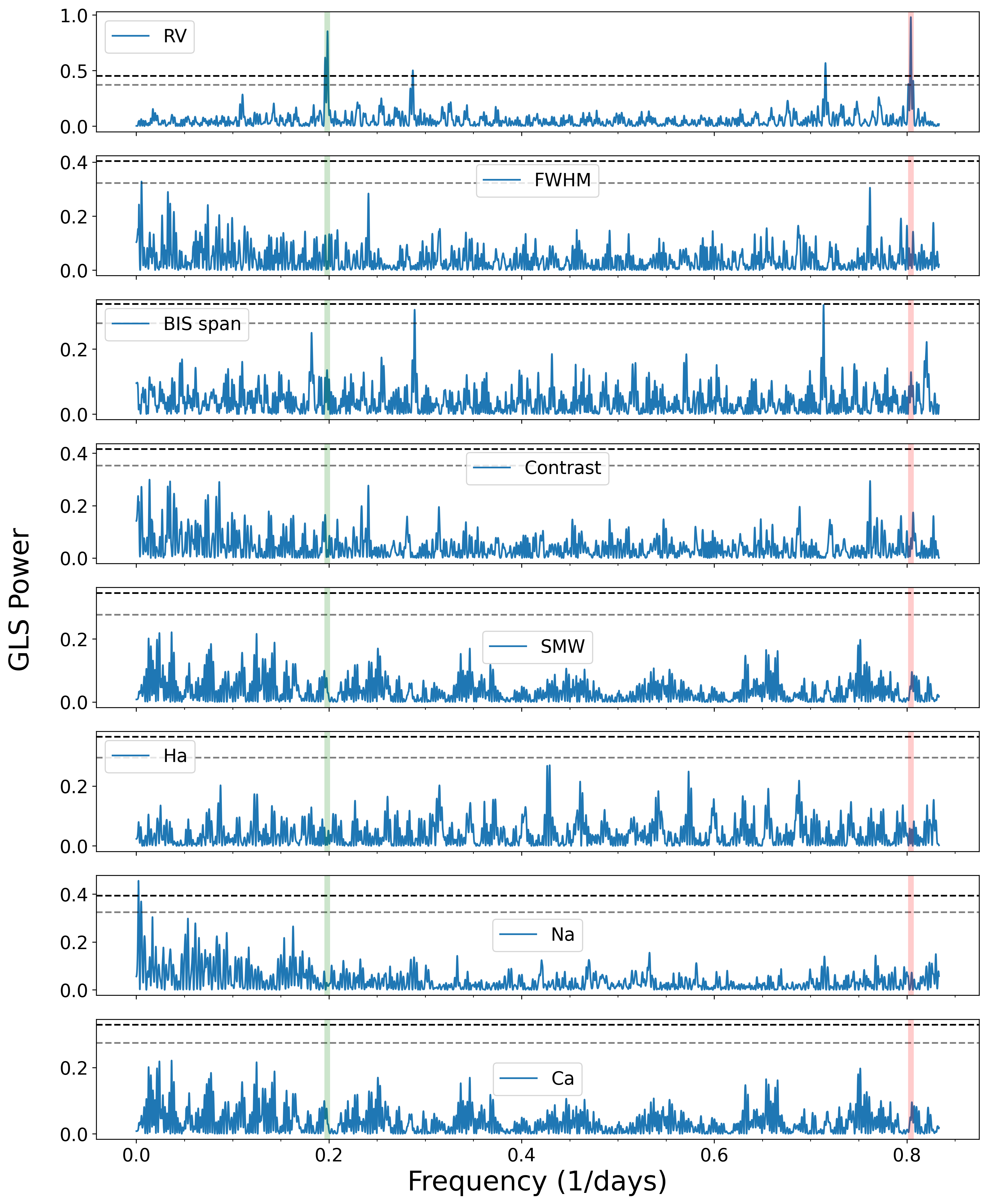}
\end{figure}
\hspace{-0.7cm} \textbf{Extended Data Figure 4: GLS periodograms of RVs and activity indexes}. The main peak of the RV GLS periodogram, at 1.24 days, and his 1-day alias are highlighted by a red and green vertical bar, respectively. The horizontal dashed lines remark the $10\%$ and $1\%$ confidence levels (evaluated with the bootstrap method), respectively.
\clearpage

\begin{figure}
\centering
\includegraphics[width=0.9\textwidth]{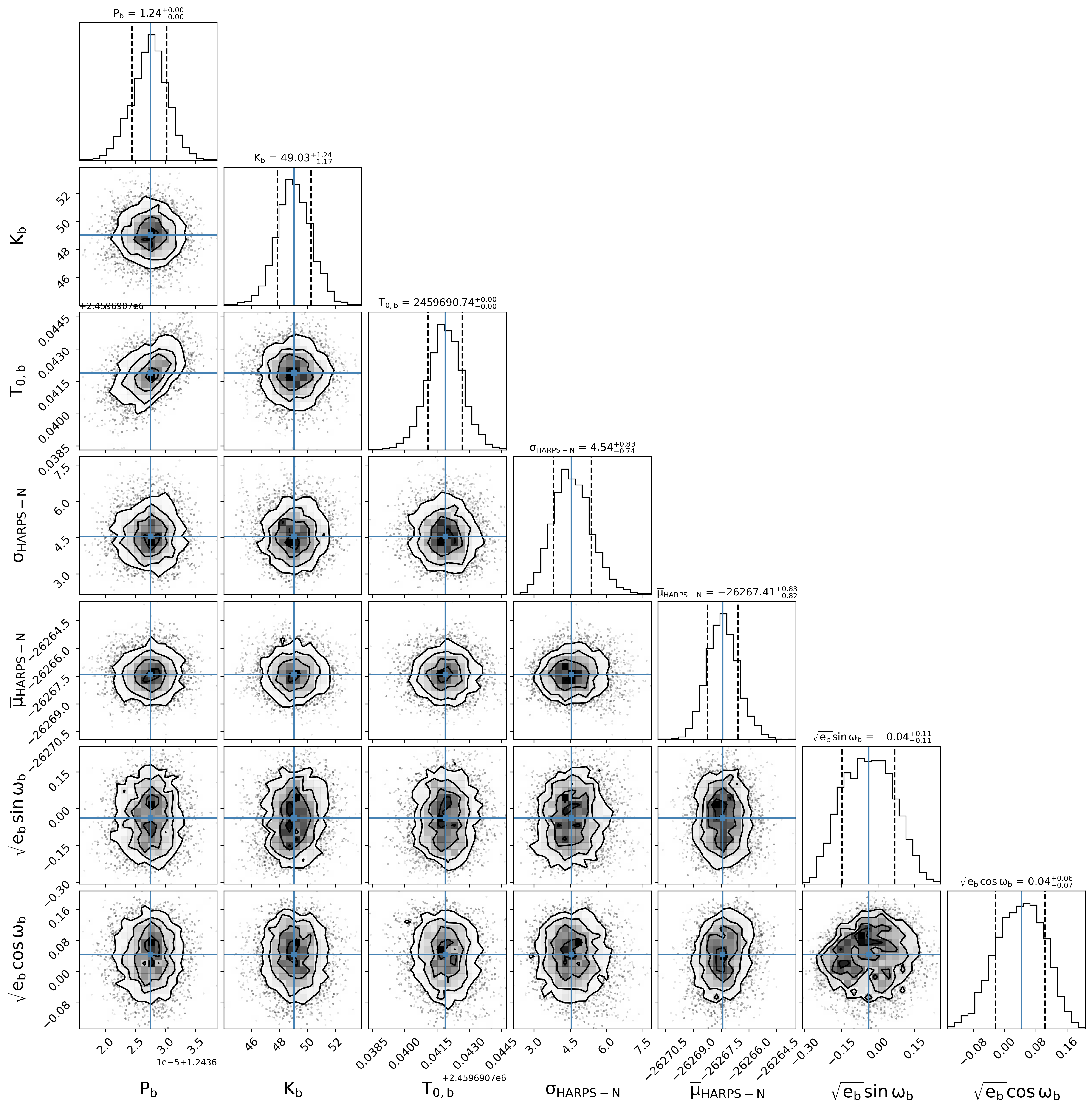}
\end{figure}
\hspace{-0.7cm} \textbf{Extended Data Figure 5: Corner plot for the posterior distributions of the global joint fit}. The blue lines indicate the average value of every parameter, while the dashed vertical lines are plotted at a distance of one standard deviation. 
\clearpage

\begin{figure}
\centering
\includegraphics[width=0.9\textwidth]{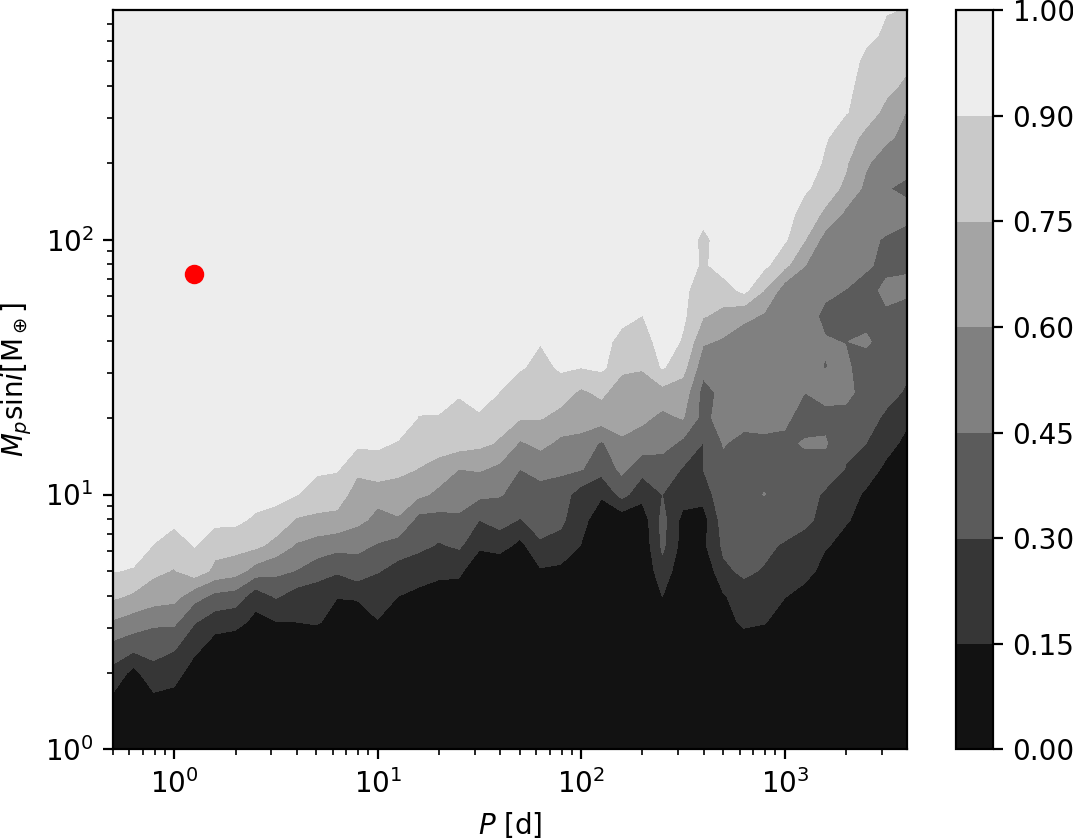}
\end{figure}
\hspace{-0.7cm} \textbf{Extended Data Figure 6: HARPS-N RV detection map}. The colour scale expresses the detection function (e.g. the detection probability), and the red circle  marks the position of TOI-1853\,b.
\clearpage

\begin{figure}
\centering
\includegraphics[width=\textwidth]{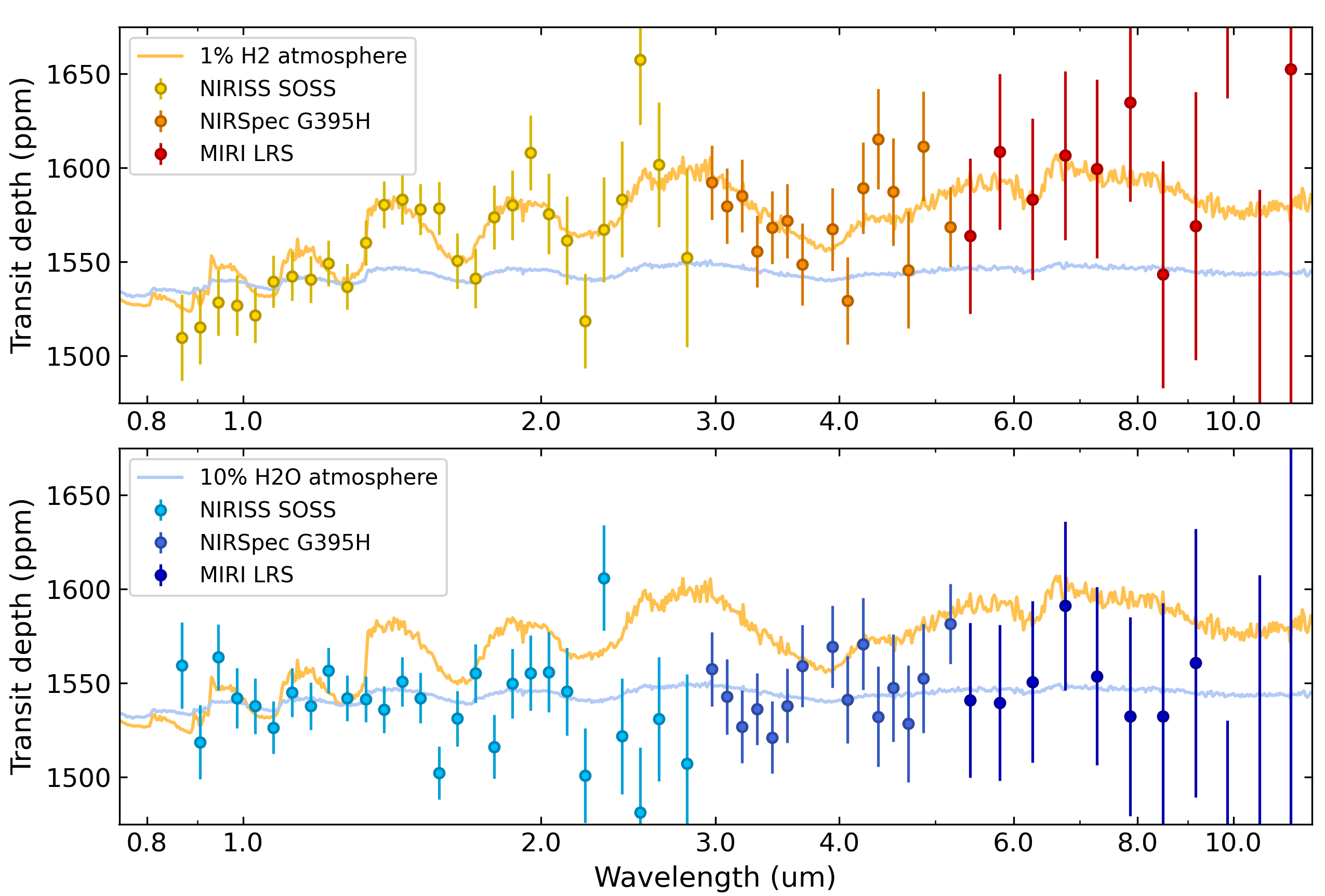}
\end{figure}
\hspace{-0.7cm} \textbf{Extended Data Figure 7: TOI-1853\,b transmission spectra simulated for JWST}. The solid curves show two synthetic spectra for an H$_2$-dominated atmosphere (orange line) and an H$_2$O-dominated atmosphere (grey line), with the respective simulated JWST observations (top and bottom panel), shown as markers with $1\sigma$ error bars, when combining 3 transits for each instrument (see labels).
\clearpage

\end{document}